\documentclass[prb,twocolumn,aps,showpacs,superscriptaddress,floatfix] {revtex4}
\usepackage[dvips]{color,graphicx}
\usepackage{dcolumn}
\usepackage{bm}
\usepackage{latexsym}
\usepackage{amsmath}
\usepackage{amssymb}
\usepackage{xcolor}

\usepackage[colorlinks,bookmarks=false,citecolor=blue,linkcolor=red,urlcolor=blue]{hyperref}

\begin{document}

\title{Frustration induced plateaux in $S \geq 1/2$ Heisenberg spin ladders}

\author{Fr\'ed\'eric Michaud}
\affiliation{Institute for Theoretical Physics, EPF Lausanne, CH-1015 Lausanne, Switzerland}
\author{Tommaso Coletta}
\affiliation{Institute for Theoretical Physics, EPF Lausanne, CH-1015 Lausanne, Switzerland}
\author{Salvatore R. Manmana}
\affiliation{Institute for Theoretical Physics, EPF Lausanne, CH-1015 Lausanne, Switzerland}
\author{Jean-David Picon}
\affiliation{Institute for Theoretical Physics, EPF Lausanne, CH-1015 Lausanne, Switzerland}
\author{Fr\'ed\'eric Mila}
\affiliation{Institute for Theoretical Physics, EPF Lausanne, CH-1015 Lausanne, Switzerland}
\date{\today}


\begin{abstract}
We study the $T = 0$ magnetization of frustrated two-leg spin ladders with arbitrary value of the spin $S$.
In the strong rung limit, we use degenerate perturbation theory to prove that frustration leads to magnetization
plateaux at fractional values of the magnetization for all spins $S$, and to determine the critical ratios
of parallel to diagonal inter-rung couplings
for the appearance of these plateaux. These ratios depend both on the plateau and on the spin. To confirm these
results, and to investigate the properties of these ladders away from the strong coupling limit, we have
performed extensive density matrix renormalization group (DMRG) calculations for $S \leq 2$.
For large enough inter-rung couplings, all plateaux simply disappear, leading to a magnetization curve typical of integer-spin chains in a magnetic field. The intermediate region turns out to be surprisingly rich however, with, upon increasing the inter-rung couplings, the development of
magnetization jumps and, in some cases, the appearance of one or more phase transitions inside a given plateau.
\end{abstract}

\maketitle

\section{Introduction}
\subsection{Motivation}
Quantum magnets in high magnetic fields possess a rich variety of physical phenomena ranging from the existence of fractional magnetization plateau\cite{kageyama1999,kodama2002} or the Bose-Einstein condensation of magnons\cite{ruegg2003} to the possible existence of the spin-equivalent of a supersolid phase.\cite{momoi2000,ng2007,sengupta2007,laflorencie2007,schmidt2008}
Of particular interest in this context are spin ladder systems.\cite{dagotto1996}
From the theoretical point of view, they constitute an interesting and non trivial step from $1D$ to $2D$.
Several materials are known to realize spin-ladders.
Most of these systems have spin $S=1/2$, but recently systems which are modeled by higher spins have been found.
An example for this is BIP-TENO, which is considered to be a frustrated $S=1$ spin ladder.\cite{sakai2004cms}

In this paper, we consider general two-leg ladders with arbitrary values of the spin $S$.
We pay particular attention to the behavior in the strong-rung limit which is amenable to an effective description in terms of perturbation theory.
Starting from this limit, we investigate the effect of frustrating interactions, in particular the appearance of additional fractional magnetization plateaux, which have already been shown to exist in several frustrated quasi-1D systems such as spin-1/2 ladders,\cite{totsuka1998,mila1998} spin-1/2 tubes,\cite{honecker2000} spin-1/2 chains with nearest and next-nearest neighbor exchange\cite{okunishi2003} and their generalization to arbitrary values of $S$\cite{HeidrichMeisner2007} as well as some models of spin-1 ladders.\cite{okazaki2002,chandra2006}
In a combined analysis using perturbative methods and the density matrix renormalization group (DMRG), we make exact predictions for the existence, the position and the sizes of these frustration induced plateaux in the strong rung limit. We then extend the analysis beyond the strong rung limit and identify an intermediate regime in which additional features, i.e., jumps in the magnetization curves and phase transitions inside the plateaux, are realized and compare to recent findings for particular values of the interactions.

Note that, since we are considering large values of the spin, one may wonder whether 
a semi-classical approach would be applicable. For the triangular lattice, the 1/3 plateau has indeed 
been shown to be an essentially classical phase of the up-up-down type stabilized in a field
range by thermal fluctuations or by quantum fluctuations treated at the level of linear spin wave theory,\cite{chubukov1991} 
and this approach has been extended to a number of other plateaux.\cite{honecker_prl} 
We believe however that such an approach cannot account for most of the plateaux
reported here for two reasons. First of all, the results depend crucially on the value of the spin
(S=1, 3/2 or 2). Besides, and more importantly, most plateaux are 'quantum' in the terminology of 
Hida and Affleck.\cite{hida} They correspond to phases that have no classical counterpart with up and 
down spins, and a semi-classical approximation in terms of fluctuations around a classical state is 
clearly inapplicable. 

\subsection{Model and Methods}
\begin{figure}[t]
\includegraphics[scale=0.7]{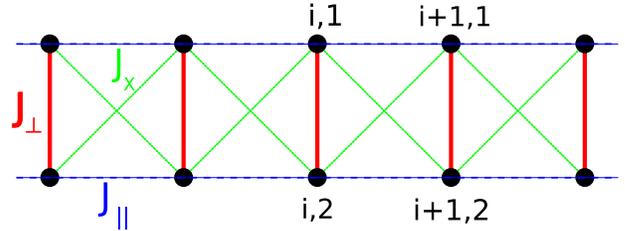}
\caption{(color online) Graphical representation of the ladder model, Eq. (\ref{eq:hamiltonian}).}
\label{ladder}
\end{figure}
We study ladder systems with frustrating interactions between the rungs depicted in Fig.~\ref{ladder}  and described by the Hamiltonian
\begin{eqnarray}
 \mathcal{H} &=& J_{\perp}\sum_i \vec{S}_{i,1} \cdot \vec{S}_{i,2} - H(S^z_{i,1} + S^z_{i,2}) \nonumber \\
 &+& J_{\parallel}\sum_i (\vec{S}_{i,1} \cdot \vec{S}_{i+1,1} + \vec{S}_{i,2}  \cdot \vec{S}_{i+1,2}) \nonumber \\
 &+& J_{\times}\sum_i (\vec{S}_{i+1,1}  \cdot \vec{S}_{i,2} + \vec{S}_{i,1}  \cdot \vec{S}_{i+1,2}) 
 \label{eq:hamiltonian}
\end{eqnarray}

Here, the $\vec{S}_{i,\alpha}$ are spin operators acting on the sites positioned on rung $i=1,\ldots,N$ and leg $\alpha=1,2$.
$J_{\perp}$, $J_{\parallel}$ and $J_{\times}$ denote the rung coupling, the inter-rung parallel coupling and the frustrating diagonal interaction, respectively, as sketched in Fig. \ref{ladder}, and $H$ is the magnetic field.
In the following, we allow the spins to be of arbitrary magnitude $S$ and we choose $\hbar = 1$.
In addition, we work in units of the energy where $J_{\perp} \equiv 1$.

Throughout the paper, the results will be discussed in terms of the magnetization per rung defined by:
\begin{equation}
M = \frac{1}{N_{\rm rung}} \sum_i \langle S_{i,1}^z + S_{i,2}^z \rangle
\end{equation}
where $N_{\rm rung}$ is the number of rungs. This magnetization varies between $0$ and $2S$, and for
isolated dimers and at zero temperature, it is a stepwise function of the magnetic field and takes
the integer values $0, 1, \ldots, 2S$.

If $J_{\perp} \gg J_{\parallel} + J_{\times}$, the behavior of the ladder is governed by the physics of single dimers and at finite fields magnetization plateaux at $0, 1, \ldots, 2S$ will appear.
In the following, we will refer to these plateaux as integer plateaux.
However, the inter-rung coupling will induce fluctuations between the dimers and in particular the competition between $J_{\parallel}$ and $J_{\times} $ will lead to new features.
If the interactions between the rungs are not too strong, these can be captured by an effective model which can be derived using degenerate perturbation theory.
This effective model in the present case is an anisotropic $S=1/2$ XXZ-chain. It can be solved exactly using the Bethe ansatz, leading to exact predictions on the positions and the sizes of possible plateaux.

The scope of the present paper is two-fold.
Starting from the strong-rung limit, we discuss the effective model.
In particular, we identify critical values of the magnetic field at which in addition to the integer plateaux new, frustration induced plateaux are created.
These predictions are compared to results obtained using the density matrix renormalization group (DMRG).\cite{white1992,white1993,schollwoeck2004,noack2005}
We consider systems with open boundary conditions (OBC) with up to $N=139$ rungs and perform, when necessary, 15 sweeps keeping maximally 1200 density matrix eigenstates.
Typically, the maximum discarded weight is of the order of $10^{-10}$.
We extrapolate our finite-size results for the size of the various plateaux to the thermodynamic limit and compare them to the predictions from the effective model for systems with spin up to $S=2$.
In order to estimate the error of our results after extrapolation, we compute numerically the gap for the XXZ-chain and perform the same extrapolation of the finite-size data as for the ladder systems.
We find that our extrapolated results for the chain agree with the Bethe ansatz up to an absolute error of the order of $5\cdot10^{-4}$. 
Assuming that the error in the extrapolation of the data for the ladder system is of a similar size, it is thus possible to compare the numerical results with a high precision to the results of the effective model.

In the second part of the paper, we leave the strong rung limit and consider cases where $J_{\parallel} + J_{\times} \geq J_{\perp}$.
For $J_{\parallel} + J_{\times} \gg J_{\perp}$, the system is described in terms of a single chain with effective spin $S_{\rm chain} = 2 S$.
Therefore, the behavior at finite fields is governed by the physics of integer spin chains.
Between the strong rung limit and the limit of an integer spin chain, we find a crossover region in which the plateaux disappear.
In this intermediate region where the description in terms of the effective model is not valid any longer, we find that the magnetization curves possess additional interesting features.
In particular, jumps in the vicinity of the plateaux are observed.
The size of the plateaux can become non-monotonic when increasing $J_{\parallel} + J_{\times}$, and it is possible to obtain phase transitions inside the plateaux as already reported in Ref. \onlinecite{chandra2006} for the special case $J_{\parallel} = J_{\times}$.

The paper is organized as follows.
In Sec. \ref{sec:strongrunglimit}, we derive the effective model and formulate its predictions for the existence and the size of additional frustration induced plateaux.
In Sec. \ref{sec:DMRGstrongrunglimit}, we present our DMRG results for ladders with $S = 1$, $S = 3/2$ and $S = 2$.
For the lowest lying frustration induced plateaux we perform a careful finite size extrapolation and compare with the quantitative predictions of the effective model.
In Sec. \ref{sec:beyond}, we leave the strong rung limit and consider the magnetization behavior for systems with $S = 1/2$ up to $S = 3/2$ as obtained by the DMRG calculations.
In Sec. \ref{sec:conclusion}, we finally summarize our findings.

\section{Frustrated ladders in the strong rung limit}

\subsection{Effective model from degenerate perturbation theory}
\label{sec:strongrunglimit}

\begin{figure}[t]
%
\includegraphics[width=0.45\textwidth]{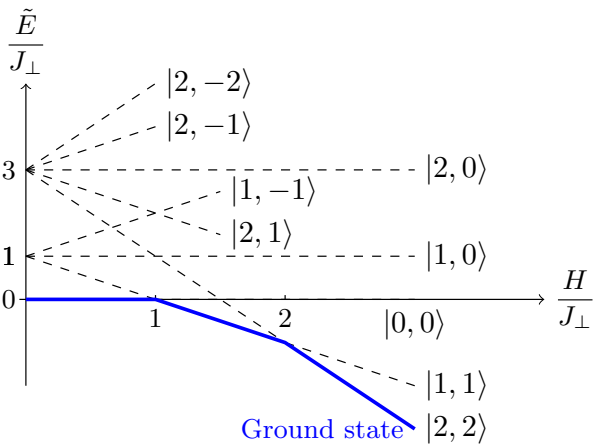}
\caption{(color online) Level crossings of a single rung for spins $S=1$. The full line denotes the ground state at different values of the magnetic field with the two critical points $H_{C_1} = J_\perp$ and $H_{C_2} = 2 J_\perp$ at which the ground state changes from singlet to triplet and from triplet to quintuplet, respectively.
Note that at the critical fields the separation between the energy levels is $J_{\perp}$, defining the energy scale for which a perturbation theoretical treatment in the strong rung limit is expected to work.}
\label{fig:single_rung}
\end{figure}

In this section we derive an effective $1D$ Hamiltonian in the strong rung limit $J_{\perp} \gg J_{\parallel}+J_{\times}$ for general values of the spin $S$. 
This limit has previously been considered in Ref.~\onlinecite{sato2005} for the case of non-frustrated ladders ($J_\times = 0$). 
Here, we consider both the inter-rung couplings $J_{\parallel}$ and the frustrating interactions $J_{\times}$ as perturbations and start from isolated dimers which can be treated exactly.
This problem has been treated in Ref.~\onlinecite{mila1998} for the case $S = 1/2$ and in Ref.~\onlinecite{sakai2004cms} for $S = 1$.
At the critical values of the fields at which the levels of the dimers cross (cf. Fig.~\ref{fig:single_rung}), we apply degenerate perturbation theory and derive an effective model for each level crossing.
(Note that  on each rung the degeneracy at all level crossings for the present Heisenberg Hamiltonian can only be two-fold).

\begin{table}[t]
	\centering
\begin{displaymath}
\begin{array}{|c|c|c|c|c|c|c|c|}
\hline
 j     & S=\dfrac{1}{2} &  S=1  & S=\dfrac{3}{2} &  S=2  & S=\dfrac{5}{2} &\ldots&  S   \\[2mm]
\hline
 1     &     1/4        &  2/3  &      5/4       &   2   &     35/12      &      & D_1(S)\\ [1mm]
 2     &     -          &  1/2  &      6/5       & 21/10 &       16/5     &      & D_2(S)\\ [1mm]
 3     &     -          &   -   &      3/4       & 12/7  &      81/28     &      & D_3(S)\\ [1mm]
 4     &     -          &   -   &       -        &   1   &      20/9      &      & D_4(S)\\ [1mm]
 5     &     -          &   -   &       -        &   -   &       5/4      &      & D_5(S)\\ [1mm]
\hline
\end{array}
\end{displaymath}
\caption{Numerical values of the first few coefficients $D_j(S)$ up to $S=5/2$ as obtained from Eq. (\ref{eq:Coeff D in terms of CG coefficients}).}
\label{tab:Djs}
\end{table}

\begin{table*}[t]
	\centering
\begin{displaymath}
\begin{array}{|c|c|c|c|c|c|c|c|}
\hline
J_{\times}/J_{\parallel}   & \rule{0pt}{3.4ex} S=\dfrac{1}{2}& S=1 &S=\dfrac{3}{2}&  S=2 & S=\dfrac{5}{2}&\ldots&  S   \\[2mm]
\hline
H_{C1}   \rule{0pt}{2.6ex} &     1/3                         &13/19&    9/11      & 15/17 &     67/73    &      & (8D_1(S)-1)/(8D_1(S)+1)\\ [1mm]
H_{C2}                     &     -                           & 3/5 &    43/53     & 79/89 &    123/133   &      & (8D_2(S)-1)/(8D_2(S)+1)\\ [1mm]
H_{C3}                     &     -                           &  -  &     5/7      & 89/103&    155/169   &      & \\ [1mm]
H_{C4}                     &     -                           &  -  &      -       &  7/9  &     151/169    &      & \\ [1mm]
H_{C5}                     &     -                           &  -  &      -       &   -   &      9/11    &      & \\ [1mm]
\vdots                     &                                 &     &              &       &              &      & \vdots \\ [1mm]
H_{C_j}                    &                                 &     &              &       &              &      & (8D_j(S)-1)/(8D_j(S)+1) \\ [1mm]
\vdots                     &                                 &     &              &       &              &      & \vdots \\ [1mm]
H_{C_{2S}}                 &     -                           &  -  &      -       &   -   &        -     &   -  & (8D_{2S}(S)-1)/(8D_{2S}(S)+1)\\[1mm]
\hline
\end{array}
\end{displaymath}
\caption{Numerical values of the critical ratios of $J_{\times}/J_{\parallel}$  beyond which frustration induced plateaux appear around fields $H_{C_j}$.}
\label{tab:SGTS}
\end{table*}

In the strong rung limit in the vicinity of the critical fields, the physics of the system is described by the two states whose energy levels cross, since the energy separation with higher energy levels is of the order of $J_{\perp}$ so that their influence can be neglected.
Thus, an effective description around the level crossings is possible by a model which takes into account two degrees of freedom on each rung.
This suggests to introduce a $S=1/2$ spin chain.
Working out the details of the degenerate perturbation theory, one indeed finds the resulting Hamiltonian to be of the form
\begin{eqnarray}\label{eq:XXZ} \nonumber
&\mathcal{H} = \sum\limits_{i}&J^{xy}(\sigma_i^x\sigma_{i+1}^x+\sigma_i^y\sigma_{i+1}^y)+J^z\sigma_{i}^z\sigma_{i+1}^z\\
&          &+\frac{H_{\textrm{eff}}}{2}(\sigma_{i}^z+\sigma_{i+1}^z),
\end{eqnarray}
where the $\sigma^{x,y,z}_i$ are the Pauli matrices acting at rung position $i$.
Thus, the resulting effective model is a $S=1/2$ XXZ-chain in a magnetic field $H_{\rm eff}$.
In the following, we will derive the parameters $J^{xy}$, $J^z$ and $H_{\rm eff}$ for the general case of spin-$S$ ladders.

\subsubsection{Mapping at the $j^{th}$ critical magnetic field}
In order to perform the perturbative treatment, we rewrite the Hamiltonian as
\begin{equation}
\begin{array}{lll}
\mathcal{H}   &=& H_0+V \\[3mm]
H_0 &=& \sum\limits_i J_{\perp} \vec{S}_{i,1}  \cdot \vec{S}_{i,2} - H_{C_j} \sum_i \left( S_{i,1}^z + S_{i,2}^z \right) \\[3mm]
V   &=& J_{\parallel} \sum\limits_i \left(\vec{S}_{i,1}  \cdot \vec{S}_{i+1,1} + \vec{S}_{i,2}  \cdot \vec{S}_{i+1,2}\right) \\
      &+& J_{\times} \sum\limits_i \left(\vec{S}_{i,2} \cdot \vec{S}_{i+1,1} + \vec{S}_{i,1}  \cdot \vec{S}_{i+1,2}\right) \\
      &-& (H-H_{C_j}) \sum\limits_i \left(S_{i,1}^z + S_{i,2}^z \right)
\end{array}
\end{equation}
Here, $H_{C_j}$ denotes the critical field at which the $j^{th}$ level crossing takes place. It is given by $H_{C_j} = j J_{\perp}$ with $j\in\{1,\ldots 2S\}$.
At these values of the fields, the ground state of the single rung is a superposition of the states $\left|j-1,m=j-1\right\rangle$ and $\left|j,m=j\right\rangle$
where the first number refers to the total spin on the rung and the second one to its projection in the $z$ direction.
Note that the ground states of $H_0$ at $H_{C_j}$ are product states of these doubly-degenerate states on the rungs.
For two adjacent interacting rungs $i$ and $i+1$, it is therefore useful to express the perturbation operator $V_i$ in the basis $\left|a\right\rangle_{i} \otimes \left|b\right\rangle_{i+1}$ with $a,b\in\{\left|j-1,m=j-1\right\rangle,\left|j,m=j\right\rangle\}$.
Since the $z$ component of the total spin is conserved under the action of $V$, the expectation value of the
operators $\vec{S}_{i,k}  \cdot \vec{S}_{i+1,l}$ ($k,l=1,2$) can be reduced to two non zero matrix elements: (i) the diagonal matrix elements
$\left\langle a\right|_i \left\langle b\right|_{i+1} S_{i,k}^zS_{i+1,l}^z \left|a\right\rangle_{i}\left|b\right\rangle_{i+1}$
which can be shown to be equal to $\frac{1}{4}\left\langle a\right|_i \left\langle b\right|_{i+1} S_{i}^zS_{i+1}^z \left|a\right\rangle_{i}\left|b\right\rangle_{i+1}$ independently of $k$ and $l$, where
$S_{i}^z := S_{i,1}^z+S_{i,2}^z$,
and (ii) the off-diagonal elements, which do not depend on $k$ and $l$ either and are given by
\begin{equation}\label{eq:Reduced Off Diagonal Element}
D_j(S) = \frac{1}{2}\left\langle j,j\right|_{i} \left\langle j-1,j-1\right|_{i+1}
                      \hat{O}
                      \left|j-1,j-1\right\rangle_{i}\left|j,j\right\rangle_{i+1}
\end{equation}
with $\hat{O}=S_{i,1}^{+}S_{i+1,1}^{-}+S_{i,1}^{-}S_{i+1,1}^{+}$.
A closed form can be obtained by expressing the rung eigenstates 
in the basis of the individual spins of the rung using Clebsch-Gordan coefficients:
\begin{eqnarray}
D_j(S) &=& \frac{1}{2} \sum\limits_{\substack{m_1+m_2=j \\ m_3+m_4=j}} \textrm{cg}(m_1,m_2;j) \, \textrm{cg}(m_3-1,m_4;j-1) \nonumber \\
&& \times \, \textrm{cg}(m_3,m_4;j) \, \textrm{cg}(m_1-1,m_2;j-1) \nonumber \\
&& \times \, \sqrt{S(S+1)-m_1(m_1-1)} \nonumber \\
&&\times \, \sqrt{S(S+1)-m_3(m_3-1)},
\label{eq:Coeff D in terms of CG coefficients}
\end{eqnarray}
where $\textrm{cg}(m_1,m_2;j)$ is the Clebsch-Gordan coefficient corresponding to the projection of the rung eigenstate $\left|j,m_1+m_2\right\rangle_i$ onto the product state of spins composing the rung $\left|S,m_1\right\rangle_{i,1} \left|S,m_2\right\rangle_{i,2}$. $m_1,m_2,m_3$ and $m_4\in\{-S,\ldots,S\}$.
The numerical values of the first few coefficients $D_j(S)$ for $S \leq 5/2$ are displayed in Tab.~\ref{tab:Djs}

At each $H_{C_j}$, the mapping to the anisotropic spin $\frac{1}{2}$ chain Eq.~(\ref{eq:XXZ}) is now achieved by introducing the pseudo-spin operators $\sigma_i$ acting on the states $\left|j-1,m=j-1\right\rangle$ and $\left|j,m=j\right\rangle$ as follows:
\begin{equation}\label{eq:Pseudo Spin}
\left\{
\begin{array}{l}
		\sigma_i^z\left|j-1,j-1\right\rangle=-\frac{1}{2}\left|j-1,j-1\right\rangle \\[2mm]
		\sigma_i^z\left|j,j\right\rangle=\frac{1}{2}\left|j,j\right\rangle \\[4mm]
		\sigma_i^+\left|j-1,j-1\right\rangle=\left|j,j\right\rangle \\[2mm]
		\sigma_i^-\left|j-1,j-1\right\rangle=0 \\[4mm]		
		\sigma_i^+\left|j,j\right\rangle=0 \\[2mm]
		\sigma_i^-\left|j,j\right\rangle=\left|j-1,j-1\right\rangle
\end{array}
\right.
\end{equation}
With this convention, the effective model takes on the form of an XXZ-chain in terms of these pseudo-spin operators. The parameters of this effective Hamiltonian can be expressed in terms of the original parameters of the ladder system as
\begin{eqnarray}
J^z &=& \frac{1}{2}\left(J_{\parallel}+J_{\times}\right) \nonumber \\
J^{xy,j}(S) &=& 4D_j(S)\left(J_{\parallel}-J_{\times}\right) \nonumber \\
H^j_{\textrm{eff}} &=& 2J^z\left(j-\frac{1}{2}\right)-\left(H-H_{C_j}\right).
\label{eq:parameters}
\end{eqnarray}
Note that $J^z$ does not depend on $j$ while the other parameters $J^{xy}$ and $H_{\rm eff}$ are different at each $H_{C_j}$.
This shows that the critical behavior, in particular the appearance and the size of plateaux, will depend on the level crossing $j$.
Note also that we expect the region of validity of the effective model to become smaller for larger spins and higher lying plateaux since the difference between the center of the plateaux defined by $H^j_{\rm eff} = 0$ and the level crossing point $H_{C_j}$ increases with $j$.

\subsubsection{Position and size of the frustration induced plateaux}

\begin{figure*}[t]
  \includegraphics[width=\textwidth]{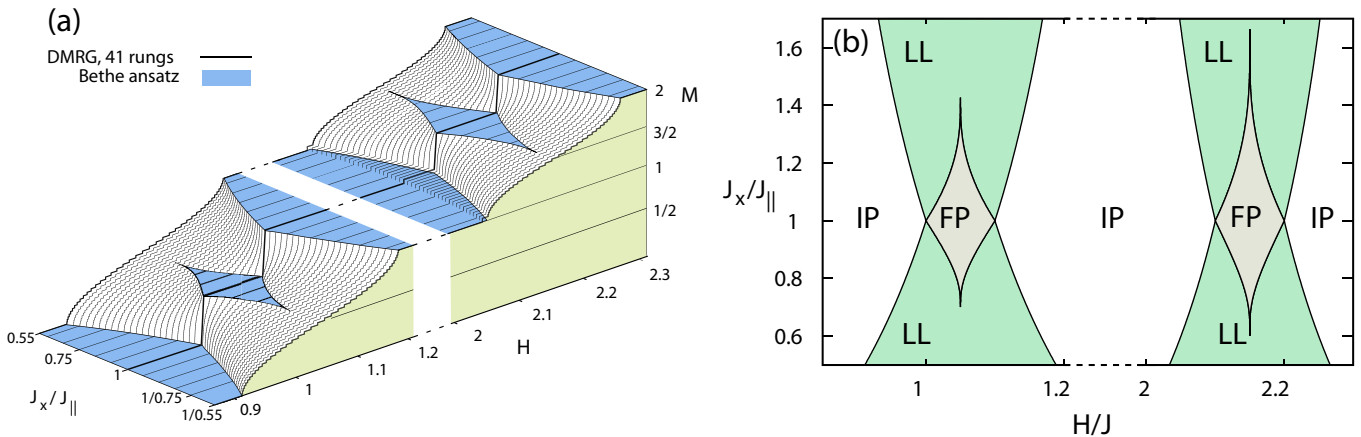}
\caption{(color online) (a) DMRG results for the magnetization of a $S=1$ ladder with $N=41$ rungs in the strong rung limit as a function of $r=J_\parallel/J_\times$ for $J_\parallel+J_\times = 0.1$.
The solid lines delimiting the plateaux are the results obtained from the Bethe ansatz solution of the effective model (\ref{eq:XXZ}) for the position and the size of the plateaux. (b) Phase diagram as obtained from the Bethe ansatz solution of the effective model. For $r=J_\parallel/J_\times \neq 1$ the frustration induced plateaux (FP) are separated from the integer plateaux (IP) by Luttinger liquid phases (LL), while for $r=1$ the plateaux are connected by jumps.}
\label{3d_spin_1_srl}
\end{figure*}

From the expressions (\ref{eq:parameters}), the Heisenberg point at which $J^z = J^{xy,j}$ translates into the condition
\begin{equation}\label{eq:Condition Heisenberg Point}
\frac{J_{\times}}{J_{\parallel}}(S)=\frac{8D_j(S)-1}{8D_j(S)+1}.
\end{equation}
As is known from the Bethe ansatz,\cite{des_cloizeaux1966} at this point a gap opens for $J^z>J^{xy,j}$, leading to a plateau in the magnetization curve of the original ladder system.
Since these plateaux are only due to the existence of frustrating inter-rung interactions, we refer to them in the following as frustration induced plateaux.
In Tab. \ref{tab:SGTS} we present the critical values of the ratios $J_{\times}/J_{\parallel}$ for general spin-$S$ ladder systems at which these plateaux appear.
Note that the Hamiltonian is symmetric under exchange of $J_{\times}$ and $J_{\parallel}$, so that at ratios larger than $1$ another critical point is obtained at which the plateaux disappear at the inverse of the ratios given in Tab. \ref{tab:SGTS}.

The size of these frustration induced plateaux can be determined by considering the analytic expression for the gap obtained from the Bethe ansatz.
For convenience, we rewrite the effective model as
\begin{equation}
\mathcal{H} = J^{xy,j}(S) (S_i^x S_{i+1}^x + S_i^y S_{i+1}^y + \rho^j(S) S_i^z S_{i+1}^z)
\end{equation}
where
\begin{equation}
\rho^j(S) = \frac{J^z}{J^{xy,j}(S)} = \frac{J_{\parallel} + J_{\times}}{8 D_j(S) (J_{\parallel} - J_{\times})}.
\end{equation}
The size of the gap for $\rho > 1$ is then found to be ($J^{xy,j}(S) \equiv 1$)\cite{des_cloizeaux1966}
\begin{equation}
\Delta(\rho) = \sinh(\Phi) \sum\limits_{-\infty}^{\infty} \frac{(-1)^n}{\cosh(n\Phi)}, \qquad \cosh(\Phi) = \rho
\label{eq:BetheGap}
\end{equation}

This function grows exponentially for $\rho \approx 1$, but becomes linear for $\rho \to \infty$.
Thus, from Eq. (\ref{eq:BetheGap}) we obtain two approximate expressions for the size of the fractional plateau $w_{\rm frac}^j$:
\begin{eqnarray}
\label{eq:sizeplateau}\nonumber
w_{\rm frac}^j &=& J_{||} + J_{\times} - 16 D_j(S) \left( J_{||} - J_{\times} \right)   \\
& & \text{   if  } \rho \gg 1 \text{, i.e., } J_{||} \approx J_{\times} \label{w} \\ \nonumber
w_{\rm frac}^j &=& 32 \pi D_j(S)\left( J_{\times} - J_{||} \right) \exp{\frac {-\pi ^2}{\sqrt{\frac{1}{D_j(S)}\left(\frac{J_{\times} + J_{||}}{J_{\times} - J_{||}}\right) -8}}}\\ & & \text{   if  } \rho \approx 1 \text{, i.e., } J_{\times}/J_{||} \approx r_c,
\end{eqnarray}

\noindent
where we denote by $r_c$ the critical value of $J_{\times}/J_{\parallel}$ at which the plateau opens.

Note that if $\Delta(\rho) > 0$, the plateau opens at the value of $H$ given by $H_{\rm eff} = 0$ in both directions when increasing and when decreasing $H$.
Thus, the size of the plateau is given by twice the size of the gap in the effective model, which has already been taken into account in the above expressions.
In addition, from the effective model using the Bethe ansatz it is possible to deduce the range over which the magnetization grows until it reaches the next integer plateau.
Thus, it is possible to obtain expressions for the size of the integer plateaux. We find
\begin{eqnarray}
\label{eq:sizeintegerplateau}
w_{\rm int}^{j,j+1} &=& J_{\perp} - 4 \left| J_{\parallel} - J_{\times}\right| (D_j(S) +D_{j+1}(S))
\end{eqnarray}
For the special case $J_{\times} = J_{\parallel}$ the size of the integer plateau is always $J_{\perp}$, as expected from isolated dimers.
The same considerations lead to expressions for the critical fields delimiting the magnetization curves between two integer plateaux. We obtain
\begin{eqnarray}
H_{c_{\rm low}} &=& J_{\perp} - 4 D_j(S) \left| J_{\parallel} - J_{\times} \right| \\
H_{c_{\rm high}} &=& 2 S\left( J_{\perp} - J_{\parallel} - J_{\times} \right) - 4 D_j(S) \left|J_{\parallel} - J_{\times} \right|
\end{eqnarray}
Finally, we would like to mention that the prediction for both the integer as well as the fractional plateaux is in full agreement with general considerations for quantum many-body systems which state that a finite excitation gap (and therefore a plateau) is only possible if $n (S-m) = {\rm integer}$, where $n$ is the periodicity of the ground state and $m$ the average magnetization per spin.\cite{OshikawaYamanakaAffleck1997,Oshikawa2000}
In the following section we will compare these predictions from the effective model to the results of our DMRG calculations.

\subsection{Frustration induced plateaux for $S = 1$, $S=3/2$ and $S=2$ }
\label{sec:DMRGstrongrunglimit}

In this section we present our DMRG results for ladder systems with $1 \leq S \leq 2$ as a function of $J_{\times}/J_{\parallel}$ keeping $J_{\times} + J_{\parallel} = 0.1$, which we expect to be sufficiently small for the systems to be in the strong rung limit.
In Fig.~\ref{3d_spin_1_srl}(a) we show our results for a finite system with $N = 41$ rungs for $S=1$.
For small values of $r = J_{\times}/J_{\parallel}$, the behavior is characterized by a very large plateau at $M = 1$ and a Luttinger liquid (LL) before and after this plateau.
For $r \approx 1$, however, two new plateaux appear, one at $M = 1/2$ and the other one at $M = 3/2$.
These are the frustration induced plateaux predicted by the effective model.
For a comparison of the predictions in the thermodynamic limit for the size and position of these additional plateaux with the finite-size DMRG results we have included the boundaries of the plateaux as obtained from Bethe ansatz.
Although the system is rather small, the comparison is excellent.
However, due to the exponentially slow opening of the plateaux at the critical fields it is very difficult to estimate the critical points in this comparison.
We will come back later to this point and present a detailed comparison after finite size scaling of the numerical results.
In Fig.~\ref{3d_spin_1_srl}(b) we present the magnetic phase diagram as obtained from Bethe ansatz.
Two remarks are in order.
First, the frustration induced plateaux are of different size and open at different critical values of $r = J_{\times}/J_{\parallel}$; the plateau at $M = 3/2$ appears at a significantly smaller value of $r$ then the plateau at $M = 1/2$.
Indeed, in order to observe the lower frustration induced plateau,
rather large values of $J_{\times}/J_{\parallel}$ are needed,
so that we expect that only strongly frustrated magnetic ladder compounds will possess this plateau.
Second, at $r=1$ the
size of the plateau as a function of $r$ has a kink.
This is due to the symmetry of the system under exchange of $J_{\times}$ and $J_{\parallel}$.

\begin{figure}[t]
\includegraphics[width=0.495\textwidth]{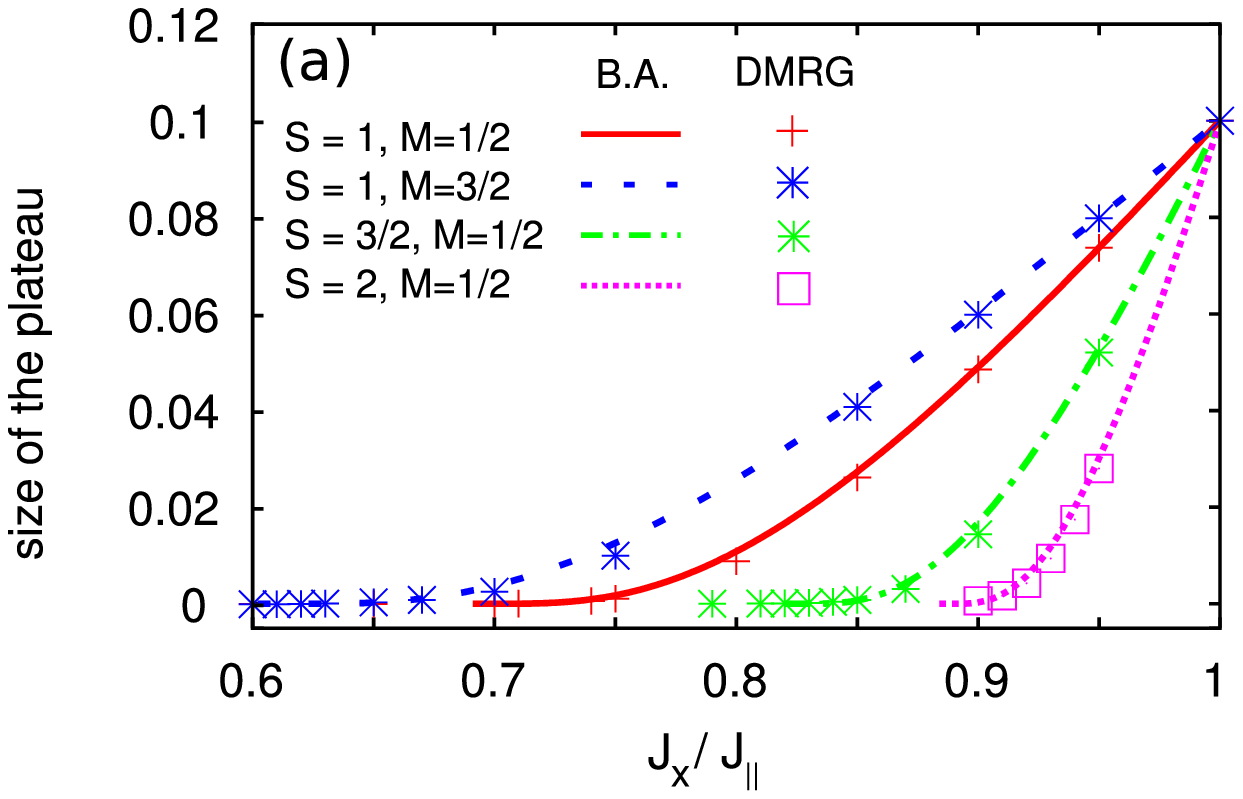}
\includegraphics[width=0.495\textwidth]{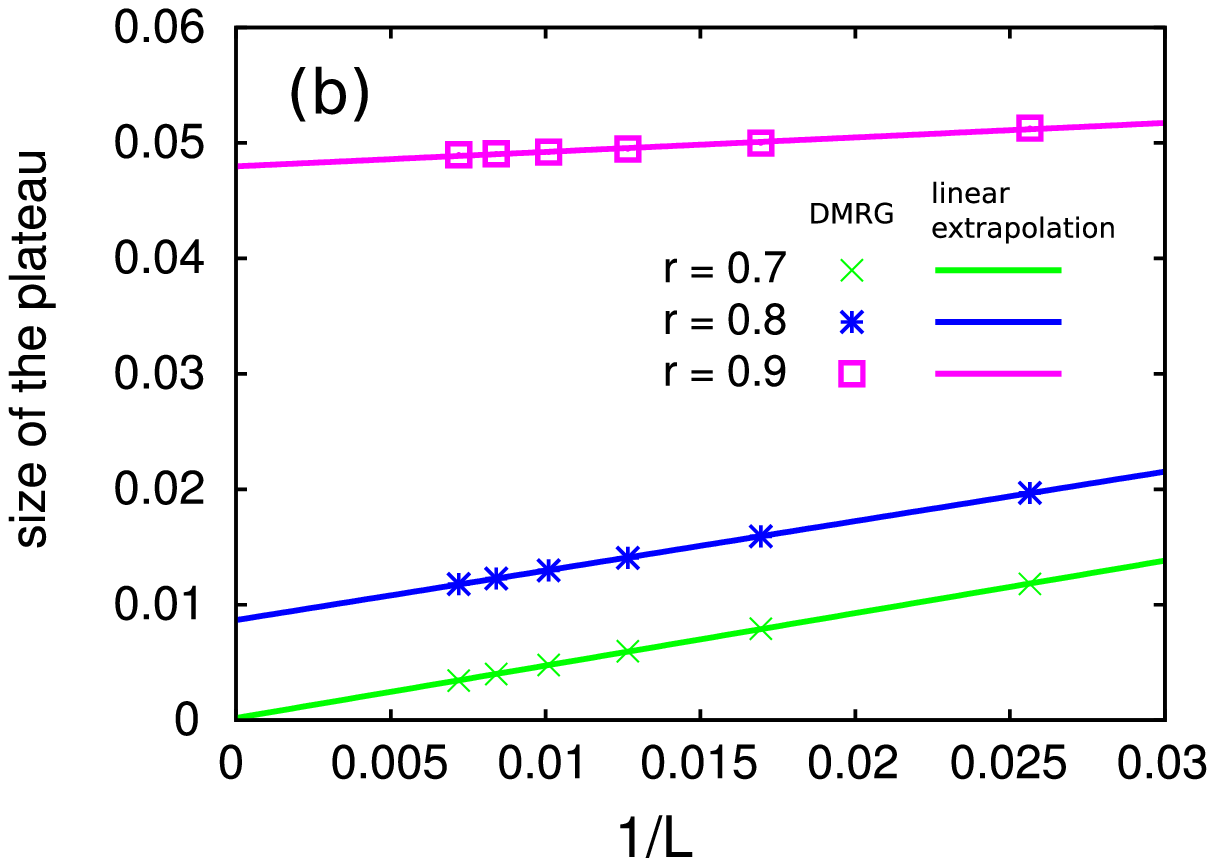}
\caption{(color online) (a) Size of some frustration induced plateaux for different values of $S$. The lines correspond to the Bethe ansatz result for the effective model and the data points are the results of finite size extrapolations of DMRG results. We estimate the error of the extrapolation to be of the order of the symbol size or smaller, see text. (b) Examples of our finite size extrapolation for some selected values of the parameters. As can be seen, a linear extrapolation is applicable for these systems.}
\label{thermodynamic_limit}
\end{figure}

Now, we turn to the finite-size extrapolation of the numerical data.
In Fig. \ref{thermodynamic_limit}(a), we present the results of this extrapolation for the size of various frustration induced plateaux as a function of $r$ for systems with $S = 1$, $S = 3/2$, and $S = 2$, and in Fig \ref{thermodynamic_limit}(b) an example of the finite size scaling for the $M = 1/2$ plateau in $S = 1$ systems is given.
Due to the symmetry under exchange of $J_{\times}$ and $J_{\parallel}$, we only discuss results for $r \leq 1$ in the following.
As can be seen, in all cases the plateau opens exponentially and possesses the same size $w_{\rm frac} = (J_{\times} + J_{\parallel})$ at $r=1$, independent of the value of $S$ in accordance to the prediction of the effective model, Eq. (\ref{eq:sizeplateau}).
The overall agreement with the Bethe ansatz results is very good, in particular for $r$ approaching 1.
However, due to the exponentially slow opening of the plateau, it is very difficult to identify numerically with a high precision the values of the critical fields, but up to this uncertainty,
they are in good agreement with the predicted values.
Note that in the parameter region where the curvature is maximal the extrapolated results show the largest deviation from the Bethe ansatz.
This deviation seems to be systematic and the obtained size of the plateau is found to be smaller than the predicted one.
This tendency becomes stronger when increasing the value of $J_{\times}+J_{\parallel}$.

Additional calculations for some selected values of the parameters for $S = 3/2$ and $S = 2$ give further support for this picture.
Since we do not expect any significant deviations from the predictions of the effective model, we refrain from presenting a detailed analysis for these cases.
We therefore conclude this section by confirming the validity of the description in terms of the effective model for values of $J_{\times}+J_{\parallel} \leq 0.1$ for $S \leq 2$ and expect it to be valid for general values of $S$.
In the next section we increase the strength of these interactions and describe our findings beyond the strong rung limit.

\begin{figure*}[t]
\includegraphics[width=1\textwidth]{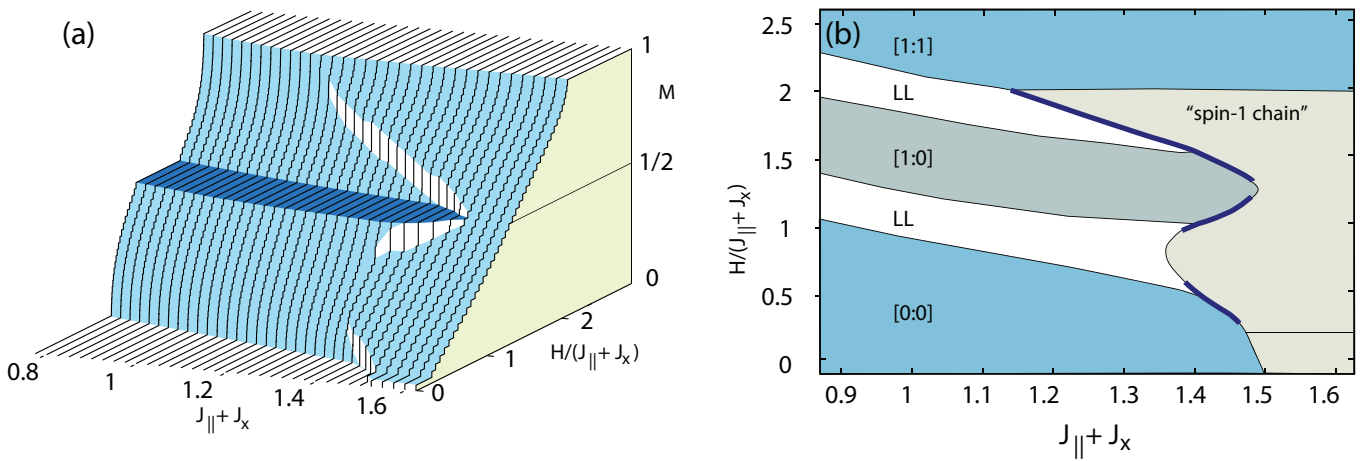}
\caption{(color online) (a) DMRG results for the magnetization of a $S=1/2$ ladder with $N=39$ rungs when changing $J_{\parallel} + J_{\times}$ and keeping  $r = J_{\parallel}/J_{\times} = 0.8$ fixed. (b) Phase diagram as obtained from the DMRG calculations for systems with $N=39$ rungs. The notation $[i:j]$ refers to the value of the total spin of two consecutive rungs $i$ and $j$ as described in the text. Thick solid lines indicate the position of jumps in the magnetization curves.}
\label{spin_12}
\end{figure*}

\section{Beyond the strong rung limit}
\label{sec:beyond}
Increasing the value of $J_{\times}+J_{\parallel}$ will eventually lead to a situation where the spacing of the energy levels of the single dimers $J_{\perp}$ is not large enough any more to consider the inter-rung couplings as perturbations.
For $J_{\times} + J_{\parallel} \gg J_{\perp}$, the system can be described in terms of a single chain with effective spin $S_{\rm eff} = 2 S$,\cite{honecker2000} leading to a continuously growing magnetization curve without any plateau.
Between the strong rung limit and the limit of integer spin chains, however, we expect to find a crossover region in which the physics is not a priori clear.
This region has been studied in Ref. \onlinecite{chandra2006} for the fully frustrated case $J_{\parallel} = J_{\times}$, and various plateaux have been identified over a wide range of $J_{\parallel}+J_{\times}$.
Interestingly, within some of these plateaux, a first order phase transition has been found to take place.
It is therefore interesting to investigate the behavior of the integer and frustration induced plateaux in the more general case $J_{\parallel} \neq J_{\times}$.
In order to keep a connection to the previous findings, we choose values of $r$ so that frustration induced plateaux exist, but which are at the same time different enough from $r=1$ so that new aspects can come into play.
Since, to our knowledge, this region of the parameter space has not yet been investigated for $S=1/2$ two-leg ladder systems, we will start our analysis with this case.
In the following, we extend it to higher spins up to $S=3/2$ in order to capture the changes in the magnetization behavior when increasing the value of $S$.
Due to the complexity of the calculations for these cases, and since the details of the transitions are not a main focus of the present work, we refrain from performing an elaborate finite size scaling analysis at this point and expect that the main features are well captured for systems of the sizes presented.
A more detailed analysis of the nature of the phase transitions and of the possible significance of finite-size effects for $S > 1/2$ ladders beyond the strong rung limit is left for future studies.

\begin{figure}[h]
\includegraphics[width=0.45\textwidth]{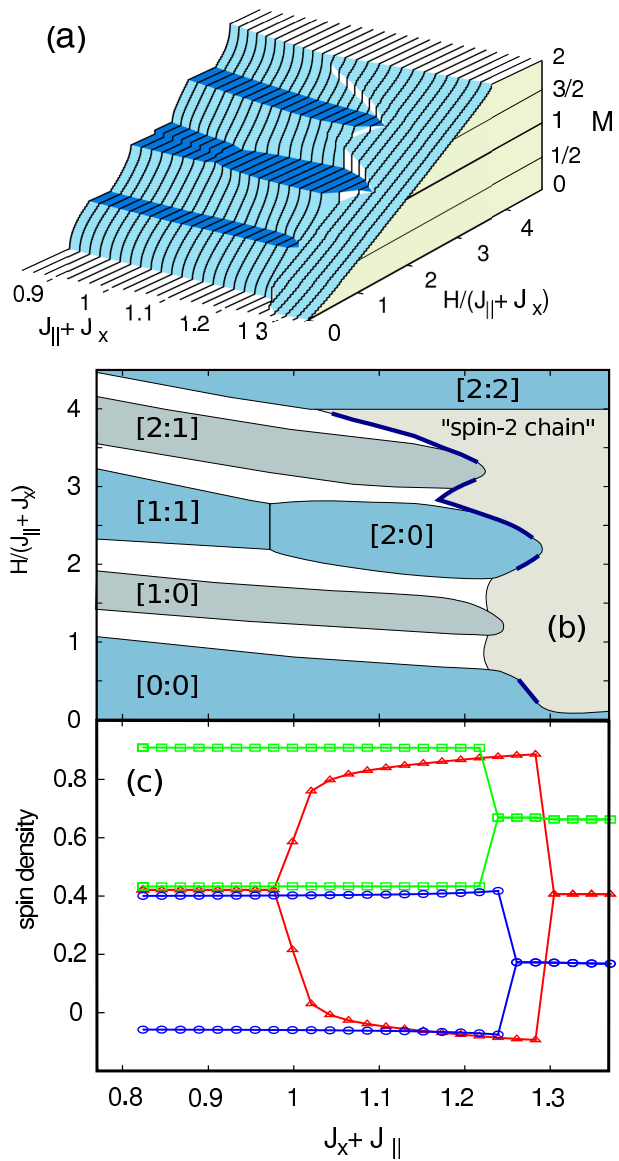}
\caption{(color online) (a) DMRG results for the magnetization of a $S=1$ ladder with $N=39$ rungs as a function of $J_{\parallel} + J_{\times}$ for $r=J_{\parallel}/J_{\times} = 0.9$. (b) Phase diagram as obtained from the DMRG results for this finite system. The notation $[i:j]$ refers to the value of the total spin of two consecutive rungs $i$ and $j$ as described in the text.  Thick solid lines indicate the position of jumps in the magnetization curves. (c) Value of $\langle S_z \rangle$ for the sites of two consecutive rungs at the center of the system as a function of $J_{\parallel} + J_{\times}$ inside the plateaux at $M = 1/2$ (\textcolor{blue}{$\bigcirc$}), $M = 1$ (\textcolor{red}{$\bigtriangleup$}) and $M = 3/2$ (\textcolor{green}{$\square$}).}
\label{3d_spin_1}
\end{figure}

In Fig.~\ref{spin_12} we present our DMRG results for $S=1/2$ ladders with $N=39$ rungs as a function of $J_{\times} + J_{\parallel}$ while keeping $J_{\times}/J_{\parallel} = 0.8$ fixed.
There are several features we would like to discuss.
Even for the rather large values of the inter-rung couplings, the fractional plateau at $M = 1/2$ exists and is of a size comparable to the one in the strong rung limit.
At $J_{\times} + J_{\parallel} \approx 1.5$, however, the plateau vanishes and the magnetization curve resembles the one of a $S=1$ spin chain.
Between $J_{\times} + J_{\parallel} \approx 1.2$ and the point at which the plateau vanishes, a jump in the magnetization curve is visible which starts at values $M \approx 1$ and comes down when increasing the value of $J_{\times} + J_{\parallel}$.
It is remarkable that this jump and the plateau both disappear around the same values of the interactions.
In addition to this jump above the fractional plateau, two other jumps appear. One reaching the $M = \frac{1}{2}$ plateau from below and the other going down to $M = 0$.
Such discontinuities indicate the position of a first order phase transition which has been discussed in Ref. \onlinecite{fouet2006} to connect a region populated by a mixture of singlets and triplets with a region consisting only of triplets in the case of the jump above the plateau.
We expect the other jumps to be of similar nature and to delimit the region in which the system gets effectively described in terms of a spin-1 chain against the region in which the ladder physics is predominant.
The plateau at $M = 1/2$ is characterized by an alternation between rungs in a singlet and a triplet state.
To describe this, we adapt the notation introduced in Ref.~\onlinecite{chandra2006} and denote possible alternating order by $[i:j]$, where the integers $i$ and $j$ denote the values of $S_{\rm total}^i = S_{i,1} + S_{i,2}$ on two consecutive rungs $i$ and $j$.
Note that for $r=1$, the rungs are in exact eigenstates of $(S^i_{\rm total})^2$, while for $r \neq 1$ this, in general, is only approximately true.
In our notation we then choose the value of $S^i_{\rm total}$ which is closest to the next integer value.

In Fig.~\ref{3d_spin_1} we present the magnetic behavior of $S=1$ ladders beyond the strong rung limit.
The overall impression is similar to the one obtained for $S=1/2$ systems.
In particular, magnetization jumps are obtained  in the vicinity of the plateaux and for $J_{\times} + J_{\parallel} > 1.3$ the magnetization curve resembles that of a spin-2 chain.
The fractional plateaux vanish slightly before the integer plateau.
For the fractional plateau at $M = 3/2$, and also for the integer plateau at $M = 1$, the point at which the plateaux disappear seems to be connected to the existence of the jumps in the magnetization curve.
An additional interesting feature is obtained for the integer plateau at $M = 1$. At $J_{\times} + J_{\parallel} \approx 1$, there is a kink in the size of the plateau.
At this point, we find indications for a phase transition taking place from a phase in which all rungs are in the triplet state to a phase with broken translational symmetry in which singlets alternate with quintuplets.
This is shown in more detail in Fig.~\ref{3d_spin_1}(c), where the local magnetizations on the two rungs at the center of the system are shown.
The plateau at $M = 1$ has a uniform spin density up to the critical value of $J_{\times} + J_{\parallel} \approx 1$ where an alternating pattern between singlets and quintuplets forms and then suddenly disappears when the plateau closes. This transition corresponds to the first-order transition found
in Ref.~\onlinecite{chandra2006} for the special case $J_{\times} = J_{\parallel}$, where at the transition the total
spin of the rungs jumps abruptly from 1 to 0 resp. 2 on alternating rungs. In the present case, despite an attempt at
a finite-size analysis, we have not been able to decide whether the transition remains
discontinuous, or whether it turns into a continuous phase transition of the Ising type, a plausible alternative
in view of the dimerized nature of the $[2:0]$ phase, and the nature of this transition as well as of the
disappearance of the plateaux is left for future investigation.
Note, however, that such a transition from a uniform to a symmetry broken state can be expected due to the fact that upon increasing the inter-rung coupling a state in which adjacent spins become more and more different is favored since it minimizes the energy of the inter-rung couplings.
For completeness, we have also shown in Fig.~\ref{3d_spin_1}(c) the magnetization pattern of the fractional plateaux.
As expected from the analysis in the strong rung limit, the fractional plateaux at $M = 1/2$ and $3/2$ have an alternating pattern between singlet and triplet rungs, or between triplet and quintuplet rungs, respectively.

\begin{figure}[t]
 \includegraphics[width=0.45\textwidth]{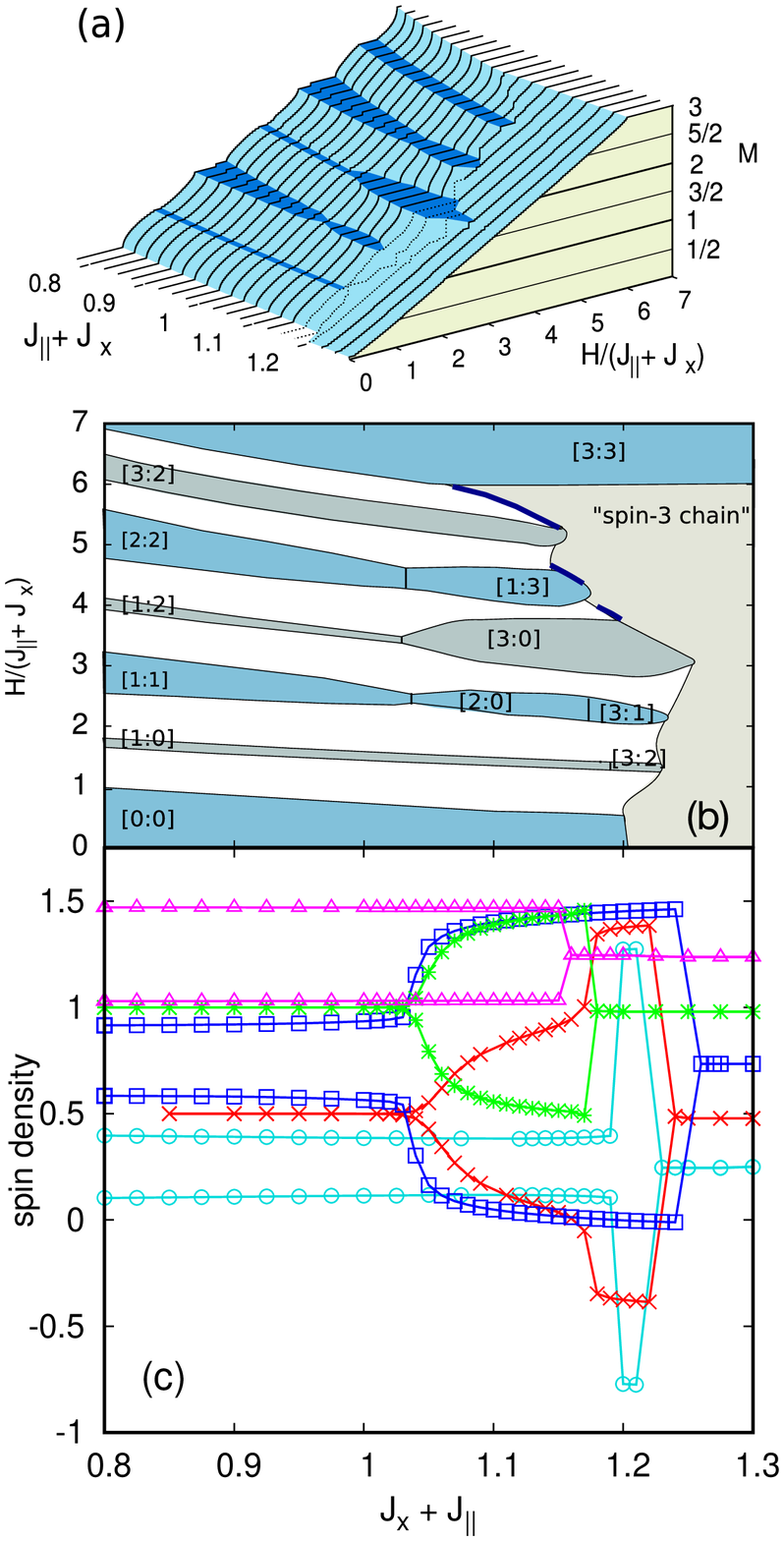}
\caption{(color online) (a) DMRG results for the magnetization of a $S=3/2$ ladder with $N=39$ rungs as a function of $J_{\parallel} + J_{\times}$ for $r=J_{\parallel}/J_{\times} = 0.9$. (b) Phase diagram as obtained from the DMRG results for these finite systems. The notation $[i:j]$ denotes the value of the total spin of two consecutive rungs $i$ and $j$ as described in the text. Thick solid lines indicate the position of jumps in the magnetization curves. (c) Value of $\langle S_z \rangle$ on the sites on two consecutive rungs at the center of the system as a function of $J_{\parallel} + J_{\times}$ inside the plateaux at $M = 1/2$ (\textcolor{cyan}{$\bigcirc$}), $M = 1$ (\textcolor{red}{$\times$}), $M = 3/2$ (\textcolor{blue}{$\square$}), $M = 2$ (\textcolor{green}{$\ast$}) and $M = 5/2$ (\textcolor{magenta}{$\bigtriangleup$}).}
\label{3d_spin_32}
\end{figure}

In Fig. \ref{3d_spin_32} finally we present our results for the magnetization curves for the $S = 3/2$ ladder systems.
The situation here is much richer than for the ladders with $S \leq 1$.
Again, for values of the inter-rung couplings large enough ($J_{\times} + J_{\parallel} > 1.25$) the magnetization curve resembles that of an integer spin chain with $S=3$ in this case.
Jumps are visible above the plateaux at $M \geq 1$.
The size of these plateaux becomes non-monotonic when increasing the values of the inter-rung interactions, indicating phase transitions similar to the one found in the $M = 1$ plateau for the $S=1$ ladder.
Note that in the region $ 1.15 < J_{\times} + J_{\parallel} < 1.25$ the lower parts of the magnetization curves show anomalous behavior, like a kink around $M \approx 0.6$.
Despite significant efforts to improve the convergence of the DMRG, in this region it turns out that the DMRG tends to get stuck at excited states and it is difficult to find the true ground states.
We depict the corresponding magnetization curves in Fig. \ref{3d_spin_32}(a) with a dotted line.
Repeated computation of these curves with much stricter convergence parameters shows that, over a wide range, the magnetization curves are well reproduced.
However, in the interesting regions the quality of the calculations remains unclear and we consider these results with caution.

In Fig. \ref{3d_spin_32}(c) we analyze the magnetization pattern inside the various plateaux.
As expected from the effective model, when the inter-rung coupling is not too large, the magnetization patterns inside the fractional plateau at $M = 1/2, \, 3/2$ and $5/2$ show a breaking of the translational symmetry and the local magnetizations alternate between values close to the ones of rung-singlets and triplets ($M = 1/2$), triplets and quintuplets ($M = 3/2$) and between quintuplets and septuplets ($M = 5/2$). As for the spin-1 case, and in agreement with the analysis of Ref.~\onlinecite{chandra2006}, the integer
plateaux undergo a transition from $[1:1]$ to $[2:0]$ and from $[2:2]$ to [1:3] respectively. However, this is not
the whole story, and Fig. \ref{3d_spin_32}(c) reveals additional and unexpected phase transitions. First of all,
phase transitions are not only found in integer plateaux, but there is clear evidence of transitions in
the $1/2$ and $3/2$ plateau. In the case of the $3/2$ plateau, it is quite similar to the tendency observed
in Ref.~\onlinecite{chandra2006}, with positive but more strongly alternating magnetizations upon increasing
the inter-rung coupling. By contrast, the transition that takes place in the $1/2$ plateau is to a ''ferrimagnetic"
state where the magnetization alternates between positive and negative values. Finally, the integer plateau at $M=1$
undergoes two transitions. After the expected transition from the $[1:1]$ phase to the $[2:0]$ phase, it
undergoes a second transition to another "ferrimagnetic"  phase with alternating positive and negative magnetization.

\begin{figure}[b]
\includegraphics[width=0.45\textwidth]{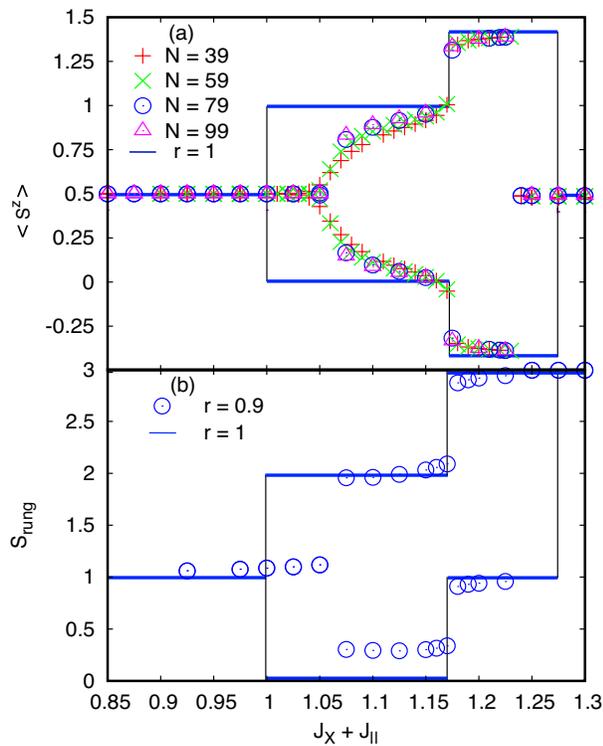}
\caption{(color online) (a) DMRG results for the magnetization pattern inside the $M = 1$ plateau for  $S = 3/2$ ladders for different system sizes when $r=0.9$ (data points) and in the thermodynamic limit for $r=1$ (continuous line).
(b) Effective value of the total spin on two adjacent rungs in the bulk in the thermodynamic limit after finite-size extrapolation for $r=0.9$ (data points) and as found from exact considerations when $r=1$ (continuous line).}
\label{sp_spin32}
\end{figure}

In order to provide further support for this picture, we show in Fig. \ref{sp_spin32}(a) the magnetization on two
consecutive rungs for different system sizes up to $N=99$ rungs for the $M = 1$ plateau.
Additional evidence is obtained by considering the value of the total spin $S^i_{\rm total}$ on two consecutive rungs in the bulk which is done by computing the expectation value $\langle (S^i_{\rm total})^2 \rangle$.
When $r=1$, we find the rungs to be in exact eigenstates of $(S^i_{\rm total})^2$ for the various system sizes under consideration.
In the case $r=0.9$, we perform a finite size scaling by considering system sizes ranging from $N=39$ up to $N=99$ rungs.
The results are shown in Fig.~\ref{sp_spin32}(b).

In the case $r=1$, we clearly identify three phases realized on the $M = 1$ plateau in which the two consecutive rungs are in eigenstates of $(S^i_{\rm total})^2$ with values of the total spin $[1:1]$, $[0:2]$, and $[1:3]$.
The first transition happens exactly at $J_{\parallel} + J_{\times} = 1$ as shown in Ref.~\onlinecite{chandra2006}.
It is a fist order transition connecting a state without broken translational symmetry with a state with broken translational symmetry.
The second transition point is obtained using numerical results for the ground state energies in the third phase, which is described by a chain of alternating spin-1 and spin-3 sites and whose ground state energy is a linear function of $J_{\parallel} + J_{\times}$.
The intersection point with the ground state energy of the system in the second phase, which is a constant since we are dealing with a product wave function of rung singlets and rung quintuplets, results in the critical point which we locate at $J_{\parallel} + J_{\times} \approx 1.17$.
This is fully consistent with the numerical results for $S_i^z$ and $S^i_{\rm total}$.


For the case $r=0.9$, after extrapolating to the thermodynamic limit, our results for $S^i_{\rm total}$ provide essentially the same picture.
However, the rungs are not in exact eigenstates of $(S^i_{\rm total})^2$, leading to non-integer effective values of the computed $S^i_{\rm total}$.
The data presented in Fig.~\ref{sp_spin32}(b) indicates the existence of a jump in $S^i_{\rm total}$ at both transitions, supporting the picture that both could be of first order.

The phase transitions to the "ferrimagnetic" configurations raise an interesting issue. First of all, we note
that no translational symmetry is broken at the transition since all phases are dimerized. But at the same time,
since the total spins of the rungs are not conserved when $r\ne 1$, these operators cannot be used as conserved
quantities to characterize the phases. It thus remains an open problem to find a way to characterize these phases.
This is reminiscent of the problem of characterizing different singlet phases in spin-1 chains with additional
interactions, or in spin-1/2 ladders. In that context, non-local string order parameters have been shown
to have different values in different phases, and to provide the appropriate way to distinguish the phases.\cite{Rommelse1989,Nishiyama1995,Fath2000}
Whether such non-local order parameters can be constructed in the present case is left for future investigation.

\section{Summary and Conclusion}
\label{sec:conclusion}
In conclusion, we have investigated the magnetic behavior of general Heisenberg spin-$S$ two-leg-ladders in a magnetic field.
Starting from the strong rung-limit, we describe the physics of the systems (independent of the value of $S$) in terms of an effective $S=1/2$ XXZ-chain obtained from degenerate perturbation theory for values of the interactions $J_{\times} + J_{\parallel} \ll J_{\perp}$.
In this limit, we predict the existence of additional fractional plateaux which are purely frustration induced and provide exact values for the position and the size of these plateaux.
Within the accuracy of our numerical resolution, we confirm these predictions with our DMRG calculations and find that the effective model for $S \leq 2$ is a qualitatively good description up to values of $J_{\times} + J_{\parallel} \approx J_{\perp}$.
In the opposite limit, for $J_{\times} + J_{\parallel} \gg J_{\perp}$, the magnetizations are reminiscent of the ones of integer $2S$ spin chains.
In an intermediate regime around $J_{\times} + J_{\parallel}  = 1.1J_{\perp} \textrm{ to } 1.5 J_{\perp}$ (depending on the actual value of $S$), we find additional interesting features in the magnetization curves.
Of particular significance are jumps in the vicinity of some of the plateaux which can also be realized at $M = 0$ and phase transitions inside the plateaux which can be visible as kinks in the size of the plateaux.
Some of these phase transitions were predicted in Ref. \onlinecite{chandra2006} for the special case $J_{\times} = J_{\parallel}$ for general values of $S$ and were found to be of first order.
In this special case, plateaux at $M = 0, \, 1/2$ and at $M = 2S, \, 2S-1/2$ are found not to possess these phase transitions, while the others should all possess at least one first order transition inside the plateaux.
Although we leave the question on the nature of the transitions open, our findings for finite systems for $S \leq 1$ are in qualitative agreement with the $r=1$ case.
However, we find indications for additional, possibly first-order transitions inside the $M = 1$ and $M = 1/2$ plateaux in the $S=3/2$ case into 'ferrimagnetic' phases where the rung magnetizations
alternate between positive and negative values. At these transitions, no symmetry is broken. Whether non-local
order parameters can be devised to characterize these phases is left for future investigation.

\section*{Acknowledgements}
We acknowledge discussions with I. Rousochatzakis and T.A. T\'oth and financial support by the Swiss National Fund and by MaNEP.



\end{document}